\documentclass[numberedrefs,preprint,turnofflinenumbers]{JASA-EL} 

\begin{document}

\title[]{Isolation performance metrics for personal sound zone reproduction systems}

\author{Yue Qiao}
\email{yqiao@princeton.edu}
\correspondingauthor
\affiliation{3D Audio and Applied Acoustics Laboratory, Princeton University, Princeton, New Jersey, 08544, USA}

\author{Léo Guadagnin}
\affiliation{FOCAL-JMlab, La Talaudière, France}
\email{l.guadagnin@focal.com}
\author{Edgar Choueiri}
\affiliation{3D Audio and Applied Acoustics Laboratory, Princeton University, Princeton, New Jersey, 08544, USA}
\email{choueiri@princeton.edu}




\begin{abstract}
Two isolation performance metrics, Inter-Zone Isolation (IZI) and Inter-Program Isolation (IPI), are introduced for evaluating Personal Sound Zone (PSZ) systems. Compared to the commonly-used Acoustic Contrast metric, IZI and IPI are generalized for multichannel audio, and quantify the isolation of sound zones and of audio programs, respectively. The two metrics are shown to be generally non-interchangeable and suitable for different scenarios, such as generating dark zones (IZI) or minimizing audio-on-audio interference (IPI). Furthermore, two examples with free-field simulations are presented and demonstrate the applications of IZI and IPI in evaluating PSZ performance in different rendering modes and PSZ robustness.
\end{abstract}


\maketitle

\textbf{The following article has been accepted by \textit{JASA Express Letters}. After it is published, it will be found at \href{https://asa.scitation.org/journal/jel}{here}.}

\section{\label{sec:1}Introduction}

Personal Sound Zone (PSZ) \citep{druyvesteyn1997personal} reproduction aims to deliver, using loudspeakers, individual audio programs to listeners in the same space with minimum interference between programs. In PSZ reproduction, given a specific audio program, a \textit{bright zone} (\textit{BZ}) refers to the area where the program is rendered for the listener, while a \textit{dark zone} (\textit{DZ}) refers to the area where the program is attenuated. For a specific listener, the audio programs are categorized into either \textit{target program}, which is rendered for the intended listener with best possible audio quality, or \textit{interfering program}, which is delivered to a different listener but may interfere with the \textit{target program} for the intended listener.


Over the past two decades, various PSZ reproduction systems have been implemented with different loudspeaker configurations and sound zone specifications depending on the application scenario. The loudspeakers used in a PSZ system have been configured as linear \citep{olivieri2013loudspeaker,okamoto2017experimental,ma2018mitigation,galvez2019dynamic}, circular \citep{chang2012sound,coleman2014acoustic,olivieri2017generation,imaizumi2021loudspeaker,baykaner2015relationship,ramo2018validating}, arc-shaped~\citep{zhu2017robust,zhu2018robust} arrays in the far field, or in the near field (e.g., headrest loudspeakers in automotive cabins\citep{elliott2006active,vindrola2021use}).The sound zones have been designed to be as large as a few meters\citep{okamoto2017experimental,olivieri2013loudspeaker,olivieri2017generation,chang2012sound,coleman2014acoustic,imaizumi2021loudspeaker}, or as small as a region that only includes the listeners' heads or ears~\citep{ma2018mitigation,galvez2019dynamic,coleman2014acoustic,zhu2017robust,zhu2018robust,elliott2006active,vindrola2021use,baykaner2015relationship}.

When evaluating the performance of PSZ systems, a commonly-adopted metric is the so-called Acoustic Contrast (AC), defined by Choi and Kim \citep{choi2002generation} as the ratio of the acoustic energy in \textit{BZ} to that in \textit{DZ}. Although the AC metric gives a measure of the separation between two sound zones, there are several limitations associated with its definition. First, AC is calculated using the sound pressure resulting from a \textit{single-channel} input that corresponds to rendering \textit{mono} audio programs. As will be shown, it does not reflect the isolation performance of other rendering modes, where \textit{multichannel} programs with different inter-channel correlations (e.g., stereo and binaural programs~\citep{canter2021delivering}) are specified as input. In addition, while AC represents the isolation between sound zones, it does not give a measure of the level difference between programs in the same zone, which is more relevant to the the listener's perception of audio-on-audio interference \citep{francombe2015model,baykaner2015relationship,ramo2018validating} (excluding psychoacoustic masking effects).

In this paper, we introduce separate PSZ performance metrics for the sound zone and audio program isolation, which are independent of both the number of the channels in a program and the correlation between channels. With these defined metrics, it is possible to evaluate the isolation performance given arbitrary program specifications from two complementary perspectives (sound zone isolation and audio program isolation). We also present examples that utilize the defined metrics to illustrate 1) the effects of different rendering modes (e.g., mono/stereo programs or binaural programs with built-in crosstalk cancellation) on the isolation performance, and 2) the effective physical boundaries of a PSZ within which the isolation performance is preserved.

The rest of the paper is organized as follows: a mathematical definition of the PSZ system is presented in Sec.~\ref{sec:2}; the new metrics are defined and discussed in Sec.~\ref{sec:3}; free-field simulations are used to illustrate two example use cases of the metrics in Sec.~\ref{sec:4}, followed by the conclusion and further discussion on the metrics and their applications in Sec.~\ref{sec:5}.


\section{\label{sec:2}Problem Definition}

\begin{figure}[h]
	\centering
	\includegraphics[width=0.6\textwidth]{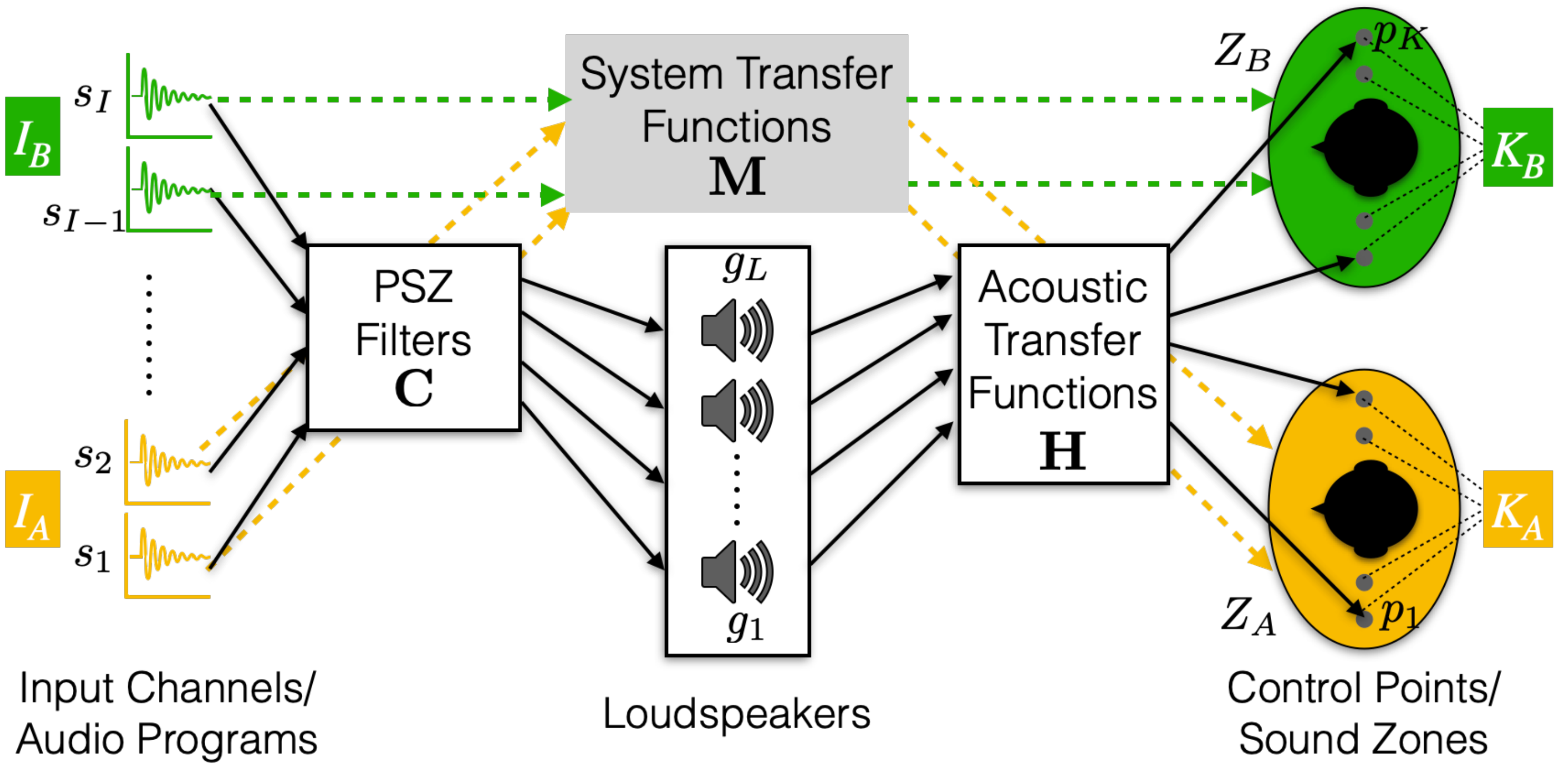}
	\caption{Signal flow diagram of a general PSZ system consisting of $I$ input audio programs (represented by virtual sources), $L$ loudspeakers and $K$ control points in two sound zones $K_A$, $K_B$. The other symbols are defined in the text below.}
	\label{fig:PSZ}
\end{figure}

Without loss of generality, we consider a PSZ system (illustrated in Fig.~\ref{fig:PSZ}) with $L$ loudspeakers and two sound zones, $Z_A$ and $Z_B$. A total of $K$ control points are defined in two zones, with the subset in each zone denoted as $\{K_A, K_B\}$, respectively. A total of $I$ channels of signals represent the inputs to the system, and are divided into two subsets corresponding to the two audio programs for the two zones, denoted as $\{I_A, I_B\}$. We use the term \textit{channel} to refer to the particular input audio signal and the term \textit{program} to refer to the unified audio content, which may contain one or more channels, as in mono or stereo/binaural programs. All subsequent quantities are implicitly dependent on frequency $\omega$; plain font is used for scalars and bold font for vectors (in lowercase) and matrices (in uppercase).

Given the complex loudspeaker gains $\bm{g}\in\mathbb{C}^{L\times 1}$, the sound pressure vector $\bm{p}\in\mathbb{C}^{K\times 1}$ at the control points is determined by 
\begin{equation}
	\bm{p} = \bm{H}\bm{g},
\label{eq:p=Hg}
\end{equation}
where $\bm{H} = \{H_{kl}\}\in\mathbb{C}^{K\times L}$ is the acoustic transfer function matrix between the loudspeakers and the control points. Furthermore, the loudspeaker gains for a channel $s_i$ of an audio program are given by
\begin{equation}
	\bm{g} = \bm{c}_i s_i,
\label{eq:g=sc}
\end{equation}
where $\bm{c}_i\in \mathbb{C}^{L\times 1}$ corresponds to the PSZ filters designed for the channel $s_i$. Combining Eq.~\ref{eq:p=Hg} and \ref{eq:g=sc}, the input channels and the control point pressure are related by
\begin{equation}
	\bm{p} = \bm{H}\bm{C}\bm{s},
\end{equation}
where $\bm{s}\in\mathbb{C}^{I\times 1}$ denotes a vector of input channels, $\bm{C}=\{C_{li}\}\in\mathbb{C}^{L\times I}$ denotes the matrix containing PSZ filters for each channel.
We further introduce the \textit{system transfer function matrix}, $\bm{M}$, as the product of transfer function and filter matrices,
\begin{equation}
	\bm{M} = \bm{H}\bm{C},
\end{equation}
where $\bm{M}=\{M_{ki}\}\in\mathbb{C}^{K\times I}$, and relates the input channels to the pressure at the control points, and will be mainly used in the subsequent metric definition. We also note that the size of $\bm{M}$ is independent of the number of loudspeakers.

\section{Metric Definition\label{sec:3}}

Two aspects of the isolation performance of PSZ systems need to be considered: the isolation between two zones given a specific program, and the isolation between target and interfering programs for a specific zone. Therefore, we define two separate isolation metrics, which are complementary for evaluating the overall system performance.

\subsection{Inter-Zone Isolation (IZI)}

To evaluate the isolation performance between the previously defined zones $Z_A$ and $Z_B$, we define the Inter-Zone Isolation (IZI) metric as the ratio of the averaged acoustic power spectra in the two zones. Since the input channels in an audio program are not necessarily correlated (for example, in a stereo program, both correlated and uncorrelated components generally exist), two terms are considered, which represent two extreme cases where all channels are correlated or none of them are correlated. IZI is then determined by the minimum of the two terms. 

The following equations show the two terms and the final IZI when $Z_A$ is considered as $BZ$:
\begin{align}
IZI_A^{corr} & = \frac{  \sum_{k\in K_A}\left|\sum_{i\in I_A}M_{ki}\right|^2/\left|K_A\right|}{\sum_{k\in K_B}|\sum_{i\in I_A}M_{ki}|^2/\left|K_B\right|} \\
IZI_A^{uncorr}& = \frac{\sum_{k \in K_A}\sum_{i\in I_A}{|M_{ki}}|^2/\left|K_A\right|}{\sum_{k\in K_B}\sum_{i\in I_A}{|M_{ki}}|^2/\left|K_B\right|} \\
IZI_A &= \min\left\{ IZI_A^{corr},IZI_A^{uncorr}\right\},
\end{align}
where $\left|K_A\right|$ and $\left|K_B\right|$ denote the number of control points in each set. \added{The definitions imply an even spacing of the defined control points, as is commonly adopted in the PSZ literature.} In the case of $Z_B$ as $BZ$, the sets $K_A$ and $K_B$ are interchanged, and $I_A$ is replaced by $I_B$.

It is worth noting that in the simplest case where only \textit{single-channel} programs are considered, IZI is equivalent to AC. Assuming the input signal to be the Dirac delta, which in the frequency domain is represented as a constant, the loudspeaker gains are given by
\begin{equation}
	\bm{q} = \bm{C}\bm{s} = \bm{c_1}s_1 = \bm{c_1},
\end{equation}
and as there is no distinction between the correlated and uncorrelated cases, IZI can be expressed as
\begin{equation}
	IZI_A = \frac{\sum_{k\in K_A}\left|M_{k1}\right|^2/\left|K_A\right|}{\sum_{k\in K_B}\left|M_{k1}\right|^2/\left|K_B\right|} = \frac{\left|\bm{H}_{A}\bm{q}\right|^2/\left|K_A\right|}{\left|\bm{H}_{B}\bm{q}\right|^2/\left|K_B\right|},
\end{equation}
where $\bm{H}_A$ and $\bm{H}_B$ are the sub-matrices of $\bm{H} $ corresponding to the two zones. The latter form is equivalent to the AC definition \citep{choi2002generation}.

\subsection{Inter-Program Isolation (IPI)}

We define the Inter-Program Isolation (IPI) metric as the ratio of the two averaged acoustic power spectra, in the same zone, corresponding to the two different audio programs. IPI therefore quantifies the isolation of the target program from the other program in a particular zone. As for the case of IZI,  we compute both the correlated and uncorrelated components of IPI, and adopt their minimum. Referring to the previously defined system, the IPI for $Z_A$ is expressed as
\begin{align}
	IPI_A^{corr} &=  \frac{\sum_{k\in K_A}\left|\sum_{i\in I_A}{M_{ki}}\right|^2/\left|I_A\right|}{\sum_{k\in K_A}|\sum_{i\in I_B}{M_{ki}}|^2/\left|I_B\right|} \\
	IPI_A^{uncorr} &=  \frac{\sum_{k \in K_A}\sum_{i\in I_A}{|M_{ki}}|^2/\left|I_A\right|}{\sum_{k\in K_A}\sum_{i\in I_B}{|M_{ki}}|^2/\left|I_B\right|} \\
	IPI_A &= \min \left\{ IPI_A^{corr},IPI_A^{uncorr}\right\},
\end{align}
\added{where $|I_A|$ and $|I_B|$ denote the number of input channels in the audio program for each zone.} For the IPI associated with $Z_B$, $I_A$ and $I_B$ are interchanged, and $K_A$ is replaced by $K_B$. 

As a special case, IPI can also be used to evaluate the isolation performance at a single control point by choosing a particular point $k$ in the set of $K_A$ (or $K_B$). We show in the subsequent section that, this single-point IPI is by itself useful in evaluating the robustness property of a generated PSZ.

\section{Applications \protect\added{Using the Pressure Matching Method}\label{sec:4}}

In this section, we show two example use cases for the complementary metrics, IZI and IPI, using sound pressure level calculated from the free-field numerical simulation of a PSZ system. \added{These examples are not able to be evaluated with the AC metric due to its limited definition.} In the first example (Sec.~\ref{sec:4.3}), IZI and IPI are used to evaluate the effects of different rendering modes on the isolation performance of the same PSZ system; in the second example (Sec.~\ref{sec:4.4}), the single-point IPI is calculated to determine the effective physical boundaries of sound zones in a two-listener PSZ system. All simulated PSZ filters in the examples are designed using the standard Pressure Matching (PM) method \citep{poletti2008investigation}. As the PM method is usually described by vectors of loudspeaker gains and pressure at control points in most of the literature, we first re-express the method in terms of filter and transfer function matrices for the case of multichannel programs.

\subsection{\protect\replaced{The reformulated Pressure Matching method}{PSZ filter generation}}

The general idea of the PM method \citep{poletti2008investigation} is to minimize the difference between the specified target sound field with ideal zone/program isolation and the actual sound field generated by the loudspeakers. The target sound field is usually specified with target pressure $\bm{p}_T$ at the control points, where in \textit{DZ} it is set to zero, and in \textit{BZ} it is set based on program-specific transfer functions. The cost function to be minimized is constructed as
\begin{equation}
	J = \left|\left| \bm{p}-\bm{p}_T \right|\right|^2 + \beta \| \bm{g} \|^2 = \left|\left| \bm{Hg}-\bm{p}_T \right|\right|^2 + \beta \| \bm{g} \|^2, 
\end{equation}
where the latter term is introduced as Tikhonov regularization to improve both matrix conditioning and the robustness of the resulting PSZ filters. Taking a single input channel to be the Dirac delta, the above function can be rewritten by replacing $\bm{g}$ and $\bm{p}_T$ with filter $\bm{c}$ and target transfer function $\bm{m}_T$, respectively:
\begin{equation}
	J = \| \bm{Hc}-\bm{m}_T \|^2 + \beta \| \bm{c} \|^2.
\label{eq:cost}
\end{equation}
The corresponding optimal solution $\bm{c}^*$ is given by 
\begin{equation}
	\bm{c}^* = (\bm{H}^H\bm{H}+\beta \bm{I})^{-1}\bm{H}^H\bm{m}_T,
\label{eq:sol_single}
\end{equation}
where $(.)^H$ denotes taking the conjugate transpose. \added{Due to added regularization, the solution applies to all three cases of $K<L$, $K=L$, and $K>L$.}  This solution is suitable only when a single-channel program is considered and the corresponding \textit{BZ} has been specified. For the case of multichannel programs, multiple targets and filters are required. Therefore, assuming $I$ channels, we modify the cost function to be the sum of the costs from minimizing the errors in all channels:
\begin{equation}
	J = \sum_{i=1}^I J_i = \sum_{i=1}^I \| \bm{H}\bm{c}_i-\bm{m}_{T,i} \|^2 + \sum_{i=1}^I \beta	\|\bm{c}_i\|^2 = \|\bm{H}\bm{C}-\bm{M}_T]\|_F^2 + \beta \|\bm{C}\|_F^2,
\end{equation}
where $\bm{C} = \left[\bm{c}_1 \cdots \bm{c}_I  \right],\bm{M}_T =  \left[\bm{m}_{T,1} \cdots \bm{m}_{T,I}  \right]$ are the filter and system transfer function matrices, respectively, and the subscript $F$ denotes the Frobenius norm. The resulting optimal filter matrix $\bm{C}^*$ is given by
\begin{equation}
	\bm{C}^* = (\bm{H}^H\bm{H}+\beta \bm{I})^{-1}\bm{H}^H\bm{M}_T,
	\label{eq:sol}
\end{equation}
which has a similar form as Eq.~\ref{eq:sol_single} for a single program except that the target vector is replaced by the target matrix. \added{Similar solutions can also be found in the literature on crosstalk cancellation systems (such as Bai and Lee~\citep{bai2006objective})}.

\subsection{Simulation setup}

\begin{figure}[h]
	\centering
	\includegraphics[width=0.47\textwidth]{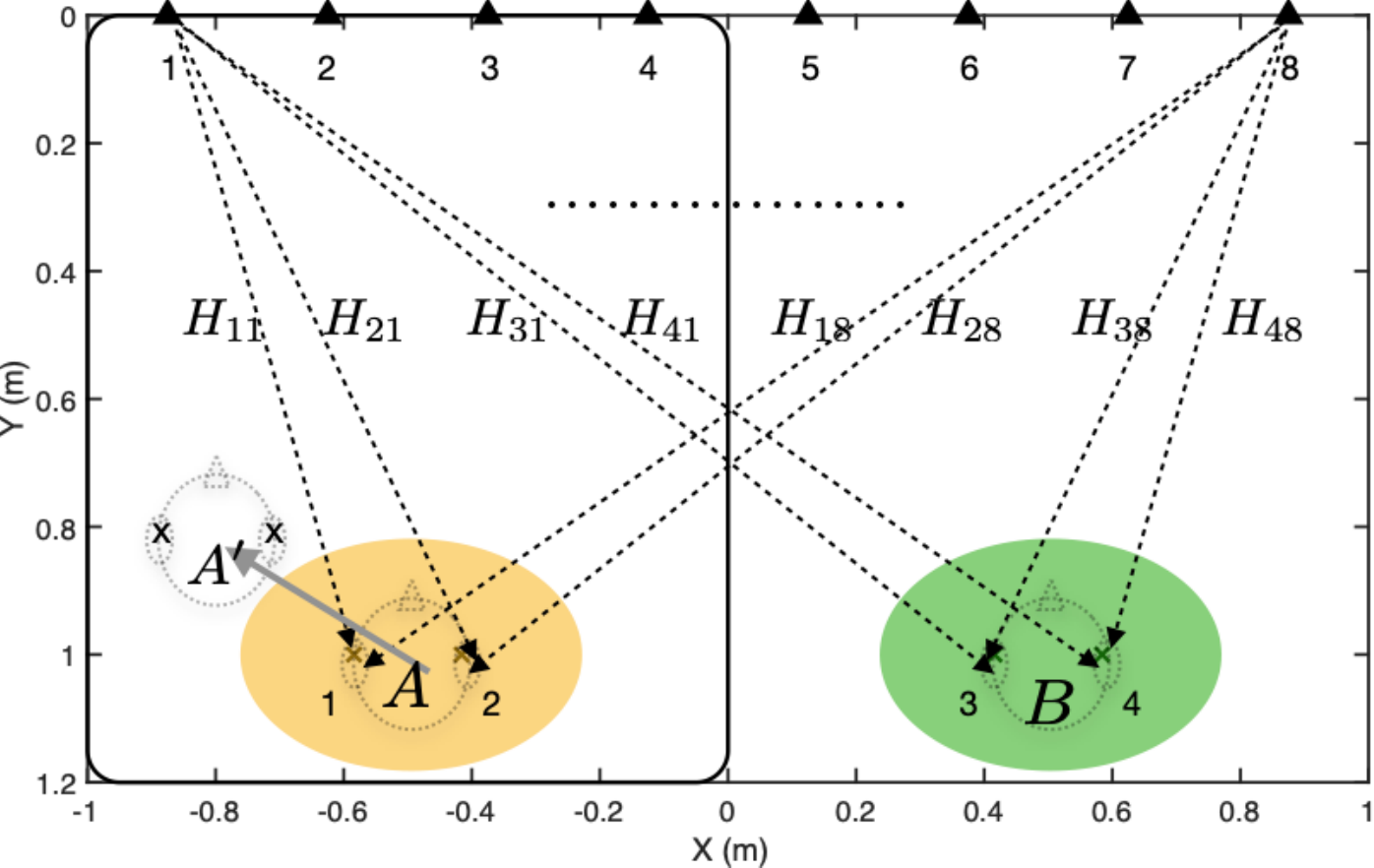}
	\caption{Illustration of the simulated PSZ system\protect\deleted{, showing the indices of loudspeakers and control points}. The black triangles on the top and the cross markers at the ear positions of two listeners represent the loudspeakers and control points, respectively, and the sound zones are indicated by two oval shadows. \protect\added{The dashed lines illustrate the indexed loudspeaker and control point for the corresponding transfer function. $A$ and $B$ represent the two centered listeners, and $A'$ represents the moved left listener.} The left half of the sound field is used to show the results in Fig.~\ref{fig:IPImap}.}
	\label{fig:system}
\end{figure}

The PSZ system adopted for the free-field simulation consists of a linear array of eight loudspeakers and two zones, as illustrated in Fig.~\ref{fig:system}. The loudspeakers are modeled as circular baffled pistons in the far field, with a spacing of 25 cm between two adjacent units. The two zones are separated by 1 m, and are also 1 m from the array. Two control points are defined in each zone, representing the ear locations of a listener with a spacing of 16.8 cm (the listeners' heads are not included in the simulation). \added{In addition to the case of both listener $A$ and $B$ located in the zone center, a moved listener (illustrated as $A'$ in the figure) with a displacement of $(x,y)=(-0.3,-0.2) m$ is also simulated.} Three cases of target matrices are considered, which correspond to three rendering modes for \textit{mono}, \textit{stereo} programs without crosstalk cancellation (XTC), and \textit{binaural} programs with XTC (equivalent to a multi-listener transaural system\citep{house2017personal}), respectively, and are given by
\begin{equation}
	\bm{M}_{mono} = \left[ \begin{array}{cc}
		\frac{H_{11}+H_{14}}{2} & 0\\ 
		\frac{H_{21}+H_{24}}{2} & 0 \\ 
		0 & \frac{H_{35}+H_{38}}{2}  \\ 
		0 & \frac{H_{45}+H_{48}}{2}
		\end{array} \right], \quad
	\bm{M}_{stereo} = \left[ \begin{array}{cccc}
		H_{11} & H_{14} & 0 & 0\\ 
		H_{21} & H_{24} & 0 & 0\\  
		0 & 0 & H_{35} & H_{38}\\ 
		0 & 0 & H_{45} & H_{48}\\ 
		\end{array} \right], \quad
	\bm{M}_{XTC} = \left[ \begin{array}{cccc}
		H_{11} & 0 & 0 & 0\\ 
		0 & H_{24} & 0 & 0\\  
		0 & 0 & H_{35} & 0\\ 
		0 & 0 & 0 & H_{48}\\ 
		\end{array} \right], \quad
\label{eq:targetM}
\end{equation}
where the transfer functions of loudspeaker 1, 4, 5 and 8 (denoted in Fig.~\ref{fig:system}) are chosen as those between the input channels and the control points. To simulate realistic uncertainties in the transfer functions due to factors such as loudspeaker position inaccuracies or response/gain variances, the \replaced{sets of transfer functions for filter design and performance evaluation are separately}{transfer functions are} sampled from a complex Gaussian distribution for each transfer function $H_{kl}$, modeled as
\begin{equation}
	H_{kl} = A_{kl} e^{i\phi_{ml}}, \quad
    A_{kl} \sim \mathit{N}(\hat{A}_{kl},\sigma_{A}^2), \quad
    \phi_{kl} \sim \mathit{N}(\hat{\phi}_{kl},\sigma_{\phi}^2),
\end{equation}
where $A,\phi$ denote the amplitude and phase of the transfer function, $\mathit{N}(\cdot,\cdot)$ denotes the normal distribution, the hat symbol denotes the value obtained from the free-field simulation, and $\sigma_A^2$, $\sigma_\phi^2$ denote the variances. In the simulation we simply choose $\sigma_A^2 = \sigma_\phi^2 = \sigma^2 =  10^{-4}$ at all frequencies, and set the constant regularization as $\beta = K\cdot\sigma^2 = 4\times 10^{-4}$ to minimize the expected cost in Eq.~\ref{eq:cost}, following a probablistic approach\citep{moller2019influence}. \added{From the same specified distribution $\mathit{N}(0,\sigma^2)$ we sample two sets of transfer functions independently for filter generation and performance evaluation, and each set is averaged across 10 trials to represent the procedure in actual experiments. The corresponding PSZ filters are computed using Eq.~\ref{eq:sol}.}

\subsection{Evaluating PSZ performance with different rendering modes\label{sec:4.3}}

In the literature, the performance of PSZ systems is usually evaluated with a single set of target pressure \citep{olivieri2013loudspeaker,chang2012sound,olivieri2017generation,vindrola2021use}, which corresponds to a fixed rendering mode for mono programs in each zone. While most existing systems are capable of delivering multichannel programs, the impact of their corresponding rendering modes on the isolation performance has not been studied. By using the new IZI and IPI metrics, such potential impact can be explicitly evaluated for various choices of target matrices.

\begin{figure}[h]
\figline{\fig{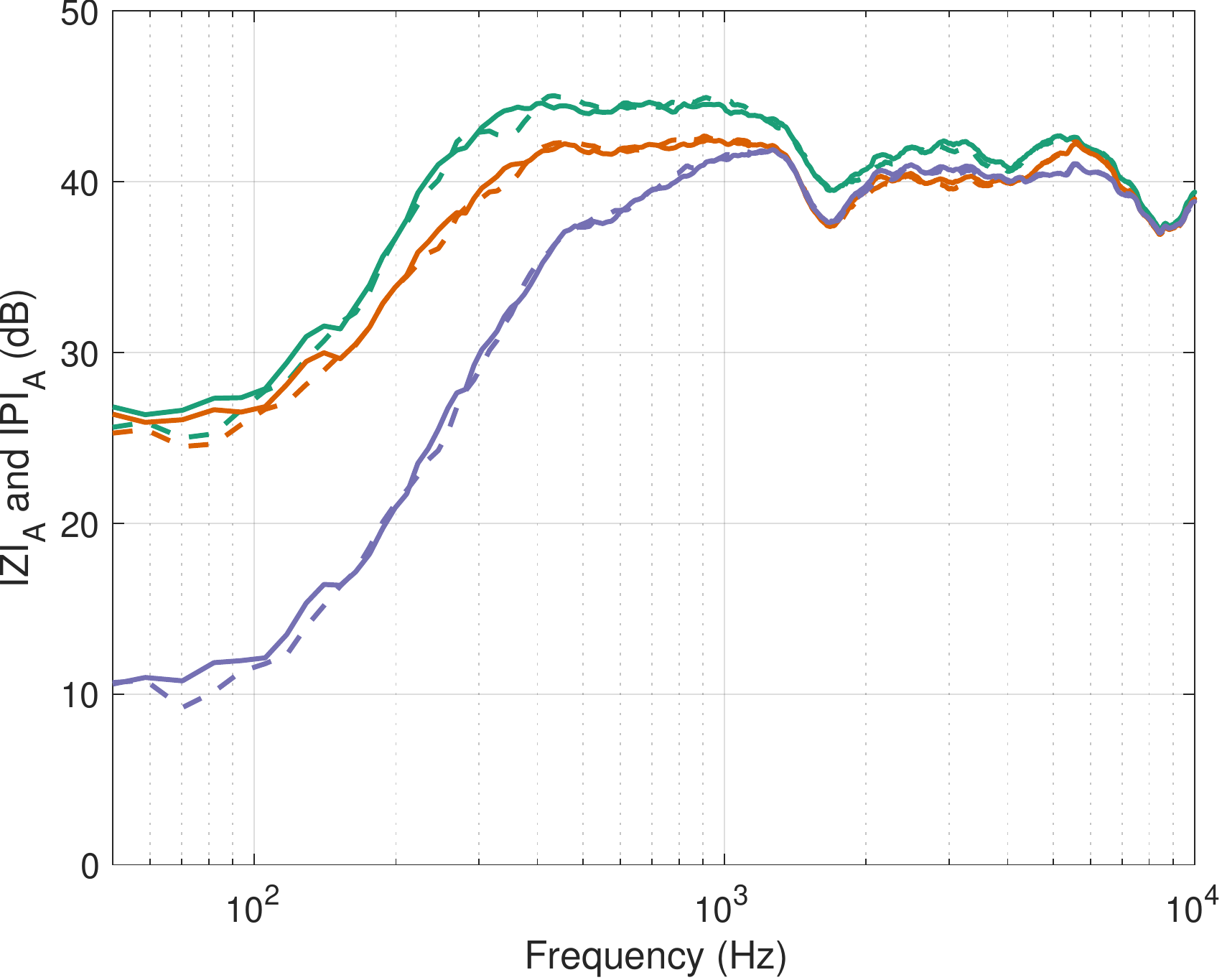}{0.35\textwidth}{(a)}\label{fig:IZIandIPI_original}
\fig{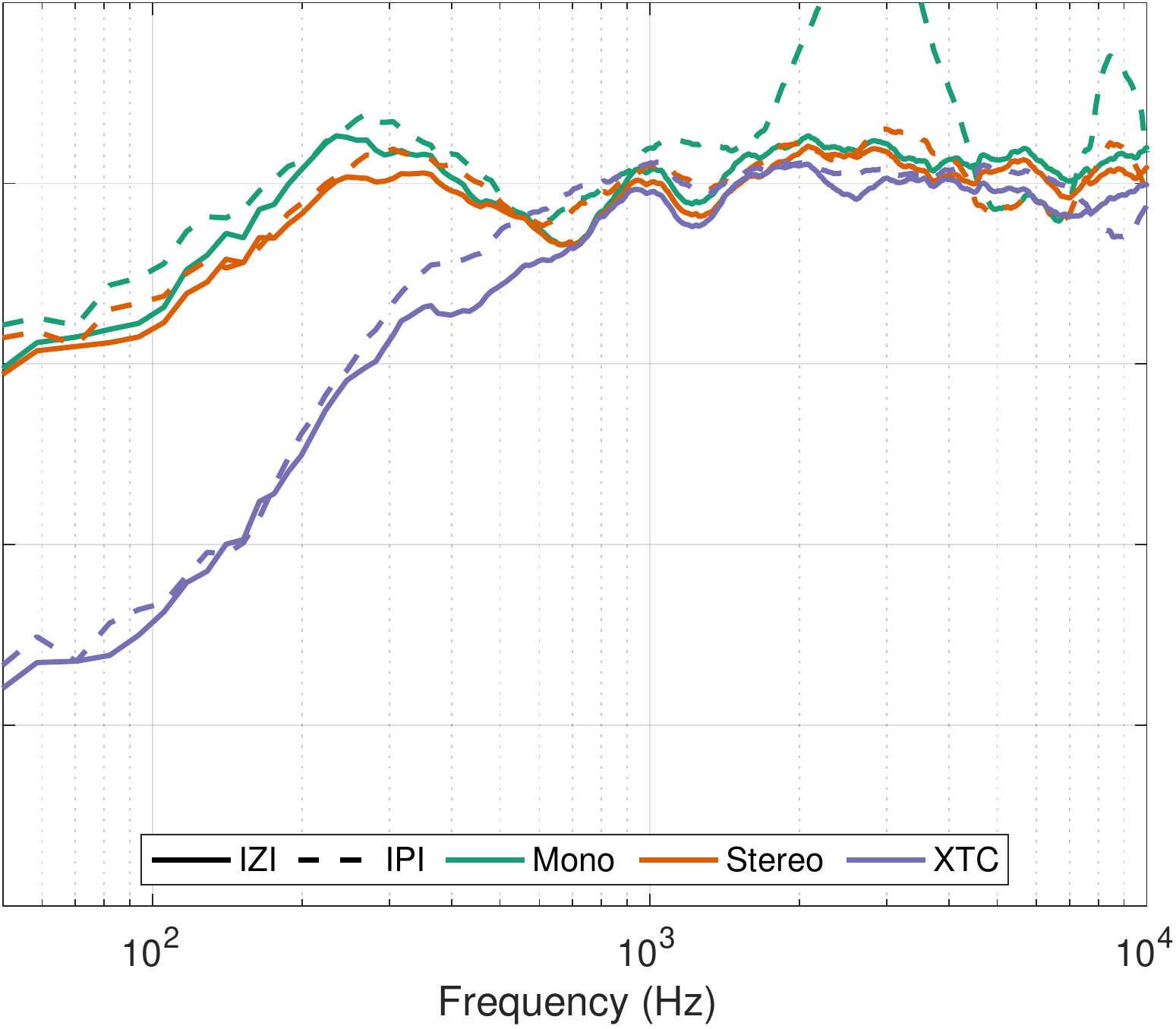}{0.32\textwidth}{(b)}\label{fig:IZIandIPI_moved}\fig{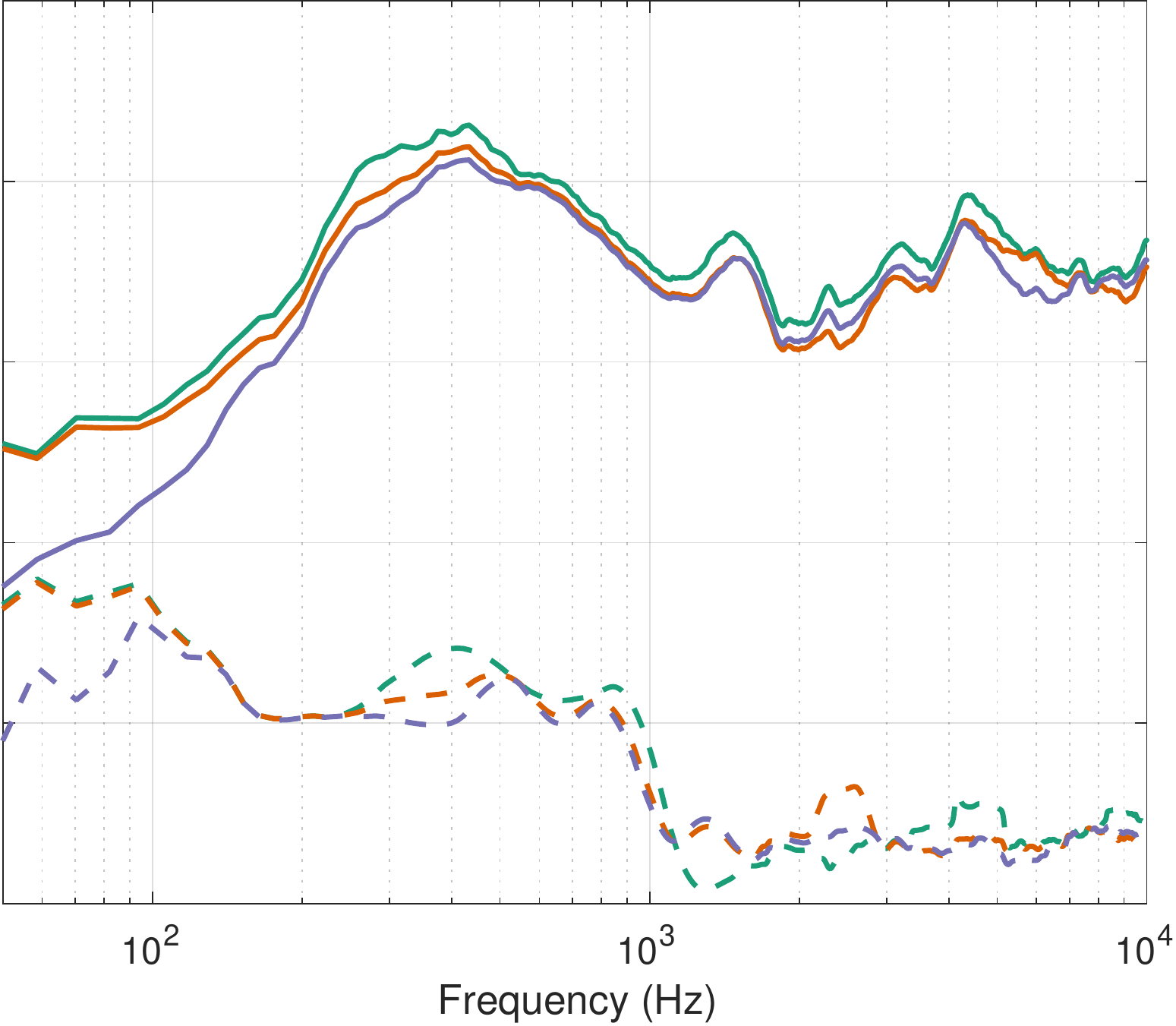}{0.32\textwidth}{(c)}\label{fig:IZIandIPI_moved_noOpt}}
\caption{\protect\replaced{Simulated IZI (a) and IPI (b) performance of the PSZ system with three different program
specifications: mono, stereo, and crosstalk-cancelled programs.}{Simulated IZI\textsubscript{A} (solid lines) and IPI\textsubscript{A} (dashed lines) for two evaluation setups of different listener positions: $\{A, B\}$ (subfigure \ref{fig:IZIandIPI_original}) and $\{A', B\}$ (subfigures \ref{fig:IZIandIPI_moved}, \ref{fig:IZIandIPI_moved_noOpt}), as illustrated in Fig.~\ref{fig:system}. The three colors represent three different program specifications: mono, stereo, and crosstalk-cancelled (XTC) programs.} 1/3-octave logarithmic smoothing is applied to the spectra for better visualization. }
\label{fig:IZIandIPI}
\end{figure}

We simulate and compare the IZI and IPI performance, of the system setup described above, for three\deleted{cases of} target matrices given by Eq.~\ref{eq:targetM} \added{and two cases of listener positions: 1) two listeners centered in both zones and 2) one listener moves away from the zone}\replaced{. Fig.~\ref{fig:IZIandIPI_original} and~\ref{fig:IZIandIPI_moved} show the simulated results respectively for IZI and IPI performance.}{, with the results for IZI\textsubscript{A} and IPI\textsubscript{A} shown in Fig.~\ref{fig:IZIandIPI}}. \replaced{Despite the introduced uncertainties, both plots present similar values and trends due to the symmetric setup and free-field assumptions.}{Fig.\ref{fig:IZIandIPI_original} corresponds to case 1, where IZI\textsubscript{A} and IPI\textsubscript{A} are almost identical due to the symmetric setup and free-field assumptions.} The \textit{mono} mode yields the best performance, followed by the \textit{stereo} mode, while the \textit{XTC} mode has the worst isolation. In particular, the degradation in the XTC mode is more significant below 1 kHz, indicating a potential trade-off between program isolation and crosstalk cancellation, due to the increased wavelength (and therefore weaker isolation between listeners) and the requirement in the cost function for cancelling the crosstalk. For PSZ systems rendering binaural audio with XTC, such trade-off can be further optimized based on established perceptual preferences of listeners \citep{canter2021delivering}, \added{or one can simply downmix the audio as mono to trade off spatial quality for better isolation. 
Fig.~\ref{fig:IZIandIPI_moved} and~\ref{fig:IZIandIPI_moved_noOpt} correspond to case 2, with the PSZ filters designed for the new and the centered position, respectively. We start to observe differences between the two metrics in~\ref{fig:IZIandIPI_moved} due to breaking of the symmetry; in~\ref{fig:IZIandIPI_moved_noOpt} the differences become clearer as the PSZ filters are no longer optimal. IZI\textsubscript{A} shows a minor degradation in isolation between two listeners, whereas IPI\textsubscript{A} reflects severely degraded isolation between two programs for the moved listener, which is more indicative of the actual experience of the listener with unoptimized PSZ filters.}

\subsection{Determining effective PSZ boundaries\label{sec:4.4}}

In most PSZ systems, sound zones are specified with regular geometries (e.g., round \citep{coleman2014acoustic,choi2002generation} or square \citep{ma2018mitigation,vindrola2021use,baykaner2015relationship,ramo2018validating} shapes). However, due to the constraints of practical systems, such as the number and distribution of loudspeakers and control points, the isolation performance within the sound zone is often non-uniform, leading to a certain lack of the robustness against possible listener movements within zones. To quantify such robustness, the effective ``boundaries'' of sound zones are defined by using the previously-defined single-point IPI metric as the contour line for a certain IPI level (e.g., 20 dB). As a result, the robustness against listener movements can be evaluated by the area (or volume) within the boundaries. Furthermore, the dependency of that robustness on moving directions can be evaluated with the projection of the effective area/volume along the direction.

\begin{figure}[h]
	\centering
	\includegraphics[width=0.8\textwidth]{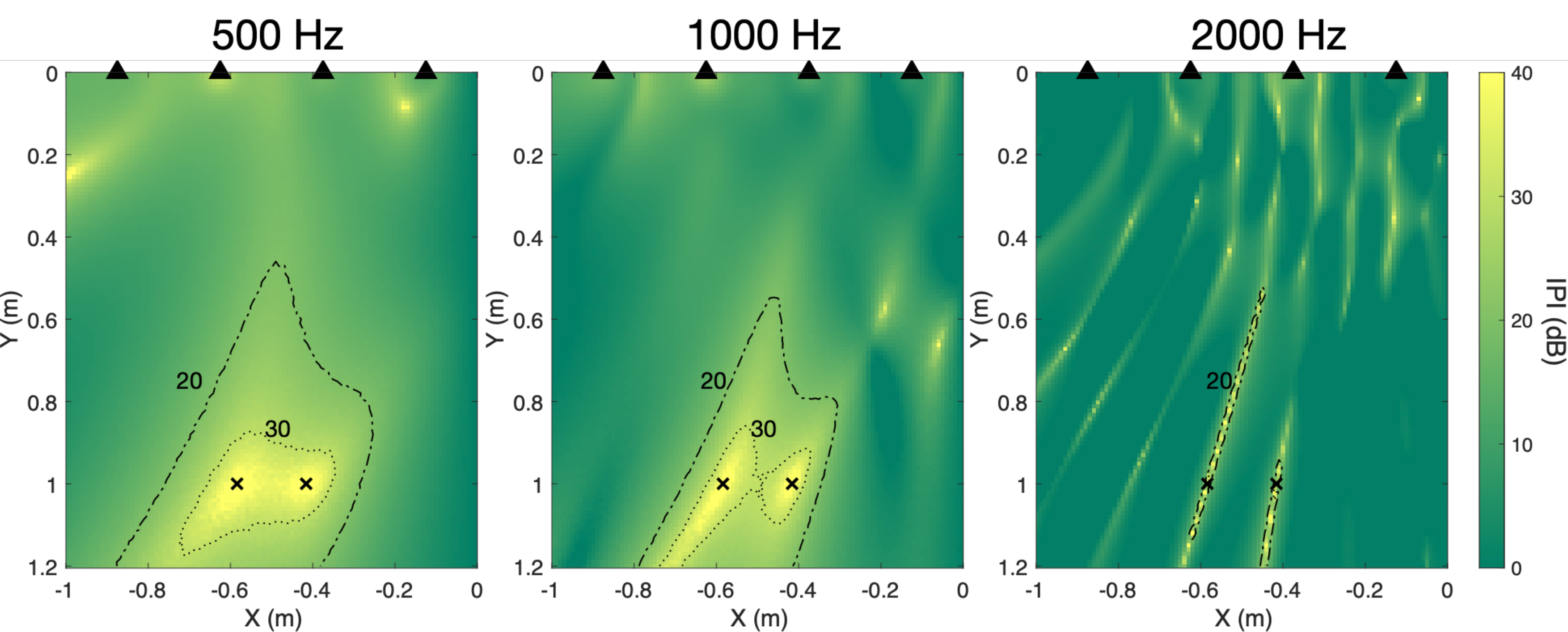}
	\caption{Computed spatial maps of single-point IPI (truncated to 40 dB) for the left half of the specified sound field, at frequencies of 0.5 (left), 1 (center), and 2 kHz (right).  Two contour lines corresponding to two IPI levels are shown in each plot with different markers (20 dB: dash-dotted line, 30 dB: dotted line).}
	\label{fig:IPImap}
\end{figure}

Fig.~\ref{fig:IPImap} shows three computed spatial maps of simulated single-point IPI in 2D for the left half of the sound field (indicated by the thick solid line in Fig.~\ref{fig:system}) at frequencies 0.5, 1, and 2 kHz, for the rendering mode of a target \textit{mono} program for the left listener. \explain{The plot for 3 kHz is replaced with that for 2 kHz as suggested by the reviewer.} In each map plot, two contour lines are shown with different line types, with the outer and inner lines corresponding to 20 and 30 dB of IPI, respectively. It is clear that given the boundaries defined by those contour lines, the shapes of the sound zones are irregular. \replaced{It can also be shown that they are highly dependent on the choice of rendering modes and system configurations.}{As PSZ filters are determined by both the exact transfer functions and the target matrices (see Eq.~\ref{eq:sol}), we also expect a great variability of the resulting PSZ boundaries with the choice of system configurations and rendering modes.} Comparing the three map plots, a trend in the decrease of the effective size of sound zones is observed as frequency increases, which has been well recognized in the literature \citep{druyvesteyn1997personal,coleman2014acoustic,vindrola2021use} \added{as reduced robustness}, but was difficult to study using the AC metric. \added{It is worth noting that the simulation 
is aimed to only illustrate the trend, which is mainly due to the wavelength changes and can be observed with and without head presence.}

\section{Conclusion and Final Discussion\label{sec:5}}

This paper introduces two metrics, IZI and IPI, for evaluating the isolation performance of generalized PSZ systems from two different aspects. The IZI metric, which reduces to the commonly-used AC metric for the special case of rendering single-channel programs, represents the isolation of the sound zones for a (single- or multi-channel) program, whereas the IPI metric quantifies the level of isolation of the target program from the interfering program in the same sound zone. 

The two metrics, albeit defined by similar expressions, are generally non-interchangeable \replaced{expect}{except} for special cases where both the physical system setup and the program assignment are perfectly symmetric with respect to the two listeners.\deleted{In reality, many factors such as room reflections and listener movements can easily break the symmetry, and therefore the two metrics should be treated separately.} In addition, the different emphases of the two metrics make them suitable for different situations: IZI compares the acoustic energy between two sound zones, and therefore is more suitable in the cases where a high contrast of sound energy between different regions is desired, such as creating a dark zone in which all audio programs are attenuated; and IPI is related to the acoustic energy of different programs rendered at the same zone, and therefore is more applicable when different programs are present concurrently, and also more suitable for the objective evaluation of the audio-on-audio interference.

In Sec.~\ref{sec:4}, we present two examples of different applications of the IZI and IPI metrics, with implications for future work. In the first example, we show the potential trade-off between the isolation performance and crosstalk cancellation at low frequencies for PSZ systems with crosstalk-cancelled binaural content. This offers the potential to further improve the filter design method by optimizing the trade-off in accordance with subjective preferences. In the second example, we show that the single-point IPI metric can serve as a basis for evaluating the robustness of the generated sound zones against listener movements. This allows the definition of a new metric that specifically quantifies the sound zone robustness, which is beyond the scope of this work and will be presented in a future study.




\begin{acknowledgments}
The authors wish to thank R. Sridhar and J. Tylka for their foundational contributions to this work. This work was supported by the research grant from the FOCAL-JMLab.
\end{acknowledgments}


\bibliography{references}

\end{document}